# Grazing incidence interaction of Sn particles with EUV Lithography ruthenium mirrors

V. Rigato

*INFN Laboratori Nazionali di Legnaro (Italy)*

## INTRODUCTION

The new EUV Lithography tools for IC High Volume Manufacture at 22nm make use of EUV radiation at $\lambda=13.5$nm. High power Laser (LPP) and Discharge (DPP) EUV light sources are based on Sn plasmas for the optimum conversion of electrical power to *in-band* radiation [1]. Sn-fueled sources emit debris such as Sn particles in a rather wide energy spectrum: from thermalized Sn to several tens keV fast ions [2,3]. Tin interaction with the collector mirrors surfaces facing the high power EUV light source leads to the degradation of the optical performance and productivity of the litho tool, therefore debris must be suppressed and the surface modification of the mirror materials during the particle irradiation must be carefully investigated both theoretically and experimentally.

For DPP Sn-fueled sources the collector is a grazing incidence mirror that reflects the EUV light in the grazing angle range from about 1° to 20°. The most used material for these collector mirrors is Ru [4].

The knowledge of the interaction process of Sn particles at different energies and angles with Ru mirrors is crucial to understand mirror degradation and to tune the parameters of the source and of the debris suppression devices to reach optimal mirror lifetime [5].

We carried out a study of the modification of the ruthenium surface exposed to the simultaneous flux of thermalized and energetic Sn at grazing incidence, for energies varying from 300eV to 30keV. The computational study is performed with the Monte Carlo codes TRIDYN [6] and TRIM.SP[7] based on binary collision approximation assumptions. These tools allow to follow dynamically and at steady state the evolution of the surface composition and to model surface binding energy and density for predicting sputtering, reflection, ion-assisted deposition and depth profiles in the nm surface region.

## MODEL ASSUMPTIONS OUTLINE

The interaction of Sn with ruthenium is studied under these initial assumptions:

1) the impinging Sn flux is composed by a thermalized flux $\Gamma_{th}$ and by a monoenergetic flux $\Gamma_{cin}$ with variable $\Gamma_{cin}/\Gamma_{th}$ ratio from 0.01 to 1.0. When $\Gamma_{cin}/\Gamma_{th}=1.0$ the standard sputtering yield of Ru by Sn ions is computed.

2) the energy of the energetic component is varied from 300eV to 30keV

3) the angle of incidence is varied from 0° (normal incidence) to 45° to check consistency with existing literature data and is then extended to the grazing incidence region ($\Theta=70°, 75°, 80°, 85°$ from normal)

4) effective surface binding energies of the surface compound species are defined by a matrix model that takes the surface composition of the target and the binding energies between the elements into account.

5) density of the $Sn_xRu_{(1-x)}$ compound is computed by using standard volume densities of the metallic species.

6) inelastic energy loss between and during collisions follow local and non-local models or equipartition of both.

7) recoil Sn and Ru atoms are followed in their motion until they stop at a cut-off energy of 0.2 eV.

The physical quantities to be studied as a function of the energy, $\Theta$, $\Gamma_{cin}/\Gamma_{th}$, and fluence $\Phi$ of Sn are 1) Sn-Sn self sputtering yield, 2) Sn-$Sn_xRu_{(1-x)}$ reflection, 3) Sn self-assisted growth rate, 4) Ru sputtering yield and erosion rate, 5) $Sn_xRu_{(1-x)}$ depth profiles and compound density.

The calculated depth profiles are then used to calculate the EUV reflectivity and to determine the degradation of the mirror [5].

## RESULTS

At grazing incidence, at low $\Gamma_{cin}/\Gamma_{th}$ values, Sn deposition prevails over Sn and Ru sputtering. A film of tin is deposited on the Ru surface with a growth rate dependent on the angle and energy.

As $\Gamma_{cin}/\Gamma_{th}$ increases, sputtering by energetic Sn becomes more and more important and eventually prevails over deposition. At a critical value of $\Gamma_{cin}/\Gamma_{th}$, the process switches from Sn self-assisted deposition to Ru sputter-erosion.

Above this value of $\Gamma_{cin}/\Gamma_{th}$, slow erosion of the Sn-Ru layer begins and a steady state surface composition is reached at a given fluence characterized by constant values of sputtering yield and stable composition.

At grazing angle Sn energetic ions are always reflected at the surface in the forward direction with high reflection coefficients (30%-99%) dependent on angle and energy.

The complexity of the Sn-Ru interaction process is better clarified in an example in Figure 1 ($E_{Sn}=1000$eV, $\Theta=85°$). Here the displacement of the surface from the original Ru surface ($\Delta x_o$-$\Delta x(\Phi)$) is computed for various $\Gamma_{cin}/\Gamma_{th}$ ratios as a function of $\Phi$ including the incorporation of Sn particles in Ru or in Sn. As it can be seen for fractions of energetic Sn ranging from 1% to about 50% the material expands while incorporating Sn particles: a Sn film is readily deposited over the Ru mirror at variable growth-rate. For fractions in the approximate range 50% to 70% the surface initially swells for few nm incorporating Sn atoms, but at some fluence the erosion of the just formed compound layer starts to dominate. As Sn bombardment continues a steady erosion of the material is achieved. For higher fractions, Ru erosion is the dominant process and readily the surface equilibrium is reached.

During Ru sputtering the incorporated Sn is concentrated in the first few nanometers from the surface. Typical calculated depth profiles of the $Sn_xRu_{(1-x)}$ compound for $E_{Sn}=1000eV$, $\Theta=85°$ are shown in figure 3 ($\Gamma_{cin}/\Gamma_{th}=0.70$): the stable surface composition is reached at fluences higher than approximately $2.5-3 \cdot 10^{16}$ $Sn/cm^2$ (see also figure 1)

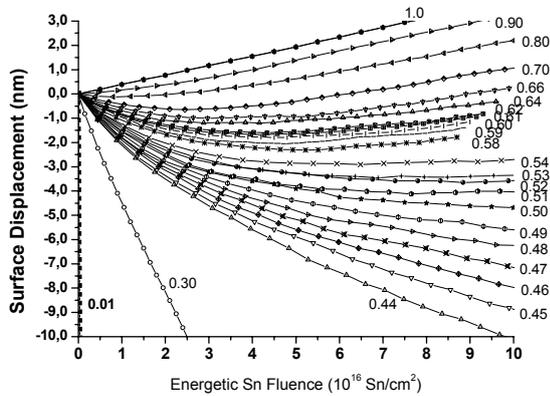

FIG. 1: *Calculated surface displacement as a function of the fluence of energetic Sn (1000eV, $\Theta=85°$) for various $\Gamma_{cin}/\Gamma_{th}$ fractions. Negative values indicate film swelling or film deposition.*

During Ru sputtering, Sn is also incorporated at the surface. In Figure 2 the areal density of the incorporated Sn is calculated for various energies at $\Theta=70°$ as a function of the fraction $\Gamma_{cin}/\Gamma_{th}$. The amount of implanted energetic Sn is of order of few $10^{15}$ $Sn/cm^2$ and is rather constant with the fraction of energetic ions in the impinging flux, while the recoiled Sn content is about one order of magnitude higher at low fractions and decreases to zero at higher fractions.

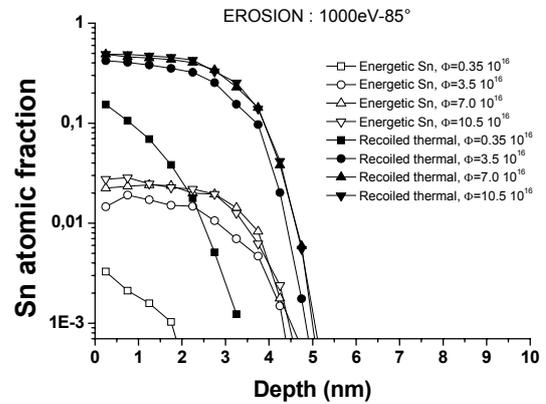

FIG. 3: *Depth profiles of implanted and recoiled Sn at various fluences during Ru sputtering ($\Gamma_{cin}/\Gamma_{th}=0.70$).*

## CONCLUSIONS

The interaction of Sn with Ru at grazing incidence is studied computationally to understand the nature of contamination of collector mirrors used in next generation EUV Lithography tools. Examples data are reported. The multi-parametric nature of the study permits to calculate the rate of deposition of Sn and of erosion of Ru for different fluxes, angles and beam spectra and to correlate the results to the contamination issues. The experimental validation of the present model requires a complex experimental laboratory setup and the exposure of the Ru mirrors to the very few high power test stands available worldwide. The film characterization will require high resolution and high sensitivity surface analysis with nuclear probes and x-rays.

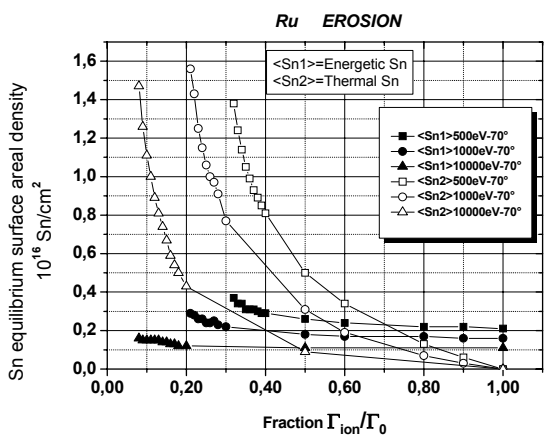

FIG. 2: *Implanted and recoiled Sn areal densities during the steady-state Ru sputtering for various angles and energies*